# MOONS Surveys of the Milky Way and its Satellites

Oscar A. Gonzalez[1]
Alessio Mucciarelli[2,3]
Livia Origlia[3]
Mathias Schultheis[4]
Elisabetta Caffau[5]
Paola Di Matteo[5]
Sofia Randich[6]
Alejandra Recio-Blanco[4]
Manuela Zoccali[7,8]
Piercarlo Bonifacio[5]
Emanuele Dalessandro[3]
Ricardo P. Schiavon[9]
Elena Pancino[6]
William Taylor[1]
Elena Valenti[10]
Álvaro Rojas-Arriagada[7,8]
Germano Sacco[6]
Katia Biazzo[11]
Michele Bellazzini[3]
Maria-Rosa L. Cioni[12]
Gisella Clementini[3]
Rodrigo Contreras Ramos[7,8]
Patrick de Laverny[4]
Chris Evans[1]
Misha Haywood[5]
Vanessa Hill[4]
Rodrigo Ibata[13]
Sara Lucatello[14]
Laura Magrini[6]
Nicolas Martin[13]
Brunella Nisini[11]
Nicoletta Sanna[6]
Michele Cirasuolo[10]
Roberto Maiolino[15]
José Afonso[16,17]
Simon Lilly[18]
Hector Flores[5]
Ernesto Oliva[6]
Stéphane Paltani[19]
Leonardo Vanzi[7]

[1] UK Astronomy Technology Centre, Edinburgh, UK
[2] Università di Bologna, Italy
[3] INAF – Astrophysics and Space Science Observatory Bologna, Italy
[4] Université Côte d'Azur, Observatoire de la Côte d'Azur, CNRS, Laboratoire Lagrange, France
[5] GEPI, Observatoire de Paris, Université PSL, CNRS, France
[6] INAF – Osservatorio Astrofisico di Arcetri, Italy
[7] Pontificia Universidad Católica de Chile, Chile
[8] Millennium Institute of Astrophysics, Chile
[9] Liverpool John Moores University, UK
[10] ESO
[11] INAF – Osservatorio Astronomico di Roma, Italy
[12] Leibniz-Institut für Astrophysik Potsdam (AIP), Germany
[13] Observatoire Astronomique, Université de Strasbourg, France
[14] INAF – Osservatorio Astronomico di Padova, Italy
[15] Department of Physics, Cavendish Laboratory, Cambridge, UK
[16] Instituto de Astrofísica e Ciências do Espaço, Universidade de Lisboa, Portugal
[17] Departamento de Física, Faculdade de Ciências, Universidade de Lisboa, Portugal
[18] ETH Zurich, Switzerland
[19] University of Geneva, Switzerland

The study of resolved stellar populations in the Milky Way and other Local Group galaxies can provide us with a fossil record of their chemo-dynamical and star-formation histories over timescales of many billions of years. In the galactic components and stellar systems of the Milky Way and its satellites, individual stars can be resolved. Therefore, they represent a unique laboratory in which to investigate the details of the processes behind the formation and evolution of the disc and dwarf/irregular galaxies. MOONS at the VLT represents a unique combination of an efficient infrared multi-object spectrograph and a large-aperture 8-m-class telescope which will sample the cool stellar populations of the dense central regions of the Milky Way and its satellites, delivering accurate radial velocities, metallicities, and other chemical abundances for several millions of stars over its lifetime (see Cirasuolo et al., p. 10). MOONS will observe up to 1000 targets across a 25-arcminute field of view in the optical and near-infrared (0.6–1.8 μm) simultaneously. A high-resolution ($R \sim 19\,700$) setting in the $H$ band has been designed for the accurate determination of stellar abundances such as alpha, light, iron-peak and neutron-capture elements.

## Scientific motivation

The need to obtain a large-scale empirical description of the Milky Way and its satellites has defined ambitious requirements for the development of cutting-edge astronomical technology during the last decade. Large international collaborations have been assembled to use this new instrumentation to produce a chemo-dynamical map of their resolved stellar populations in exquisite detail. However, there are several questions that unfortunately cannot be fully addressed with the available instrumentation.

The central regions of the Milky Way, specifically those near the Galactic plane, are currently inaccessible to high spectral resolution observations with sufficiently large number statistics and spatial coverage because of an inability to efficiently overcome the effect of the immense amounts of interstellar dust across the disc. As a consequence, we are currently unable to reconstruct the entire history of the Milky Way down to its most central components, thus missing one of the critical pieces in the puzzle of galaxy formation. The only way to remedy this is to carry out a coherent, large-scale mapping of the stellar populations of the Milky Way, particularly focused on the in-plane regions of the bulge and the inner disc, providing the key missing ingredient necessary to fully understand the processes behind the formation and evolution of barred disc galaxies.

Similarly, the globular cluster system of the Milky Way has long been used to learn about the evolution of the Galaxy. The study of the stellar populations of globular clusters in the inner Milky Way can give us important clues about the early stages of the Galaxy's formation by tracing the most vigorous epochs of in situ star formation as well as its early accretion history. However, the study of inner, highly reddened Milky Way clusters is hampered by the small sample of stars with suitable infrared spectroscopy and the lack of homogeneity with respect to field stars. Constructing a large homogeneous sample of stars from inner globular clusters and the surrounding fields with high-quality measurements of kinematics and chemistry — which can be homogeneously compared to those of the bulge — is critical to understanding the origin of the clusters and their contribution to the relatively metal-poor component of the Milky Way bulge.

On the other hand, the study of the formation and evolution of young star clusters and their stellar populations represents the backbone of research in modern astrophysics, as it has a strong impact on our understanding of key open issues, from the processes of star and planet formation to the assembly and evolution of the Milky Way and galaxies in general. However, a homogeneous determination of precise radial velocities and chemical content of the optically faint cluster populations (intrinsically redder objects, embedded stars, embedded/reddened clusters currently not accessible because of their large extinctions and densities) is vital to achieving full sampling of the parameter space (for example, lower masses, presence of gas, position in the Milky Way disc).

Furthermore, beyond the Milky Way, the proximity of the galaxies in the Local Group allows us to observe their individual stars both with spectroscopy and photometry, as well as to study in detail the characteristics of their stellar populations. In particular, the closest satellites of the Milky Way are the Large Magellanic Cloud (LMC), the Small Magellanic Cloud (SMC), and the remnant of the Sagittarius dwarf spheroidal galaxy. These three galaxies are excellent targets with which to investigate and unravel the star formation history of irregular/dwarf galaxies that have experienced gravitational interactions between each other and/or with the Milky Way. Thus, chemical and kinematical information about the stellar populations in the Magellanic Clouds and the Sagittarius dwarf spheroidal galaxy is a fundamental complement to the corresponding information for the Milky Way.

The MOONS surveys will be focused on the detailed kinematic and chemical characterisation of the resolved stellar populations in the Milky Way and its closest satellites, with a focus on environments that continue to be poorly sampled by previous and ongoing spectroscopic surveys. We will dedicate two main surveys to this objective:

– Reddened Milky Way Survey (70 nights), which will sample the chemo-dynamics of the stellar populations in the Galaxy's inner 3 kpc and in a few other disc regions affected by severe extinction in the visual;

– Milky Way Satellites Survey (30 nights), which will sample the stellar populations in the central disc/bar region of the Magellanic Clouds and in the Sagittarius galaxy and streams.

Both surveys will obtain high-signal-to-noise spectra ($SNR \sim 50–100$) to obtain temperatures and surface gravities (from photometric and spectroscopic diagnostics), abundances of iron, $\alpha$-elements (from optical and near-infrared spectra), CNO and possibly a few other light elements (from near-infrared spectra only), and radial velocities.

### The reddened Milky Way survey

The MOONS REDdened Milky WAY (REDWAY) survey will sample 0.5 million stars across the Galactic plane (Figure 1) tracing red-clump stars in the bulge/bar and inner disc regions, red-giant stars of the inner globular cluster system and the nuclear bulge, as well as mapping young embedded star clusters. It will take full advantage of three aspects of MOONS: i) its high-density targeting capabilities to obtain up to 900 stars in each 25-arcminute diameter pointing and the remaining ~ 100 fibres on sky regions[a], ii) its high-resolution mode — in $RI$ ($R \sim 9200$), $YJ$ ($R \sim 4300$), and $H$ ($R \sim 19\,700$) — to disentangle key spectral features, and iii) the 8-m aperture of the VLT and high efficiency of MOONS to produce exquisite, high-signal-to-noise spectra in the near-infrared.

### The nuclear region

A key ingredient for understanding the processes taking place in the central regions of galaxies is to perform resolved stellar population studies in our own Galaxy. MOONS will provide the opportunity to obtain high-resolution spectra for stars in the nuclear star cluster and disc, thus allowing the measurement of radial velocities, [Fe/H], CNO, alpha, and a few other elements.

The survey will cover the nuclear bulge with 15 fields (blue circles in Figure 1) covering $-1 < l < 1$ degrees and $-0.5 < b < 0.5$ degrees corresponding to a scale height coverage of ~ 50 pc and scale length of ~ 140 pc. A total of ~ 13 000 red giant stars will be observed. In the central field, at $l,b = 0,0$ degrees, a subset of 200 stars will sample the innermost 10 pc region, therefore mapping the nuclear star cluster.

The resolved stellar populations and kinematics of these extreme systems, as seen in the Milky Way by MOONS, will become a unique benchmark template for unresolved nuclear clusters in external galaxies as well as providing vital input regarding their formation scenarios and establishing their link to their larger-scale host components.

### The bulge/bar at low latitudes

In order to understand the complex structure (and the formation mechanism) of components potentially co-existing in the in-plane regions of the Galaxy (for example, Wegg, Gerhard & Portail, 2015), it is important to obtain a large sample of stars mapped across all of the mid-plane to characterise and identify the orbits and chemical abundances of stars. In order to satisfy this objective, a suitable grid of fields (shown in Figure 1 in orange) has been designed to provide the best coverage and number statistics to trace the properties of the different Galactic components to an unprecedented level of detail. The highest priority in terms of time investment has been given to the mid-plane region of the bulge where the field coverage is nearly contiguous. This is a unique property of this survey, which is not possible with any other facility because these high-density stellar fields are heavily reddened.

The survey will obtain ~ 900 red-clump stars across each of the 380 fields (at a SNR > 60 in $H$ band) in these poorly explored regions, allowing us to derive statistically significant distributions of metallicities, alpha-element abundances (and potentially ages), as well as 3D velocities by combining proper motions from the VISTA Variables in the Via Lactea (VVV) survey with MOONS radial velocities to distinguish between different families of orbits and possible independ-ent components that can be compared with state-of-the-art simulations.

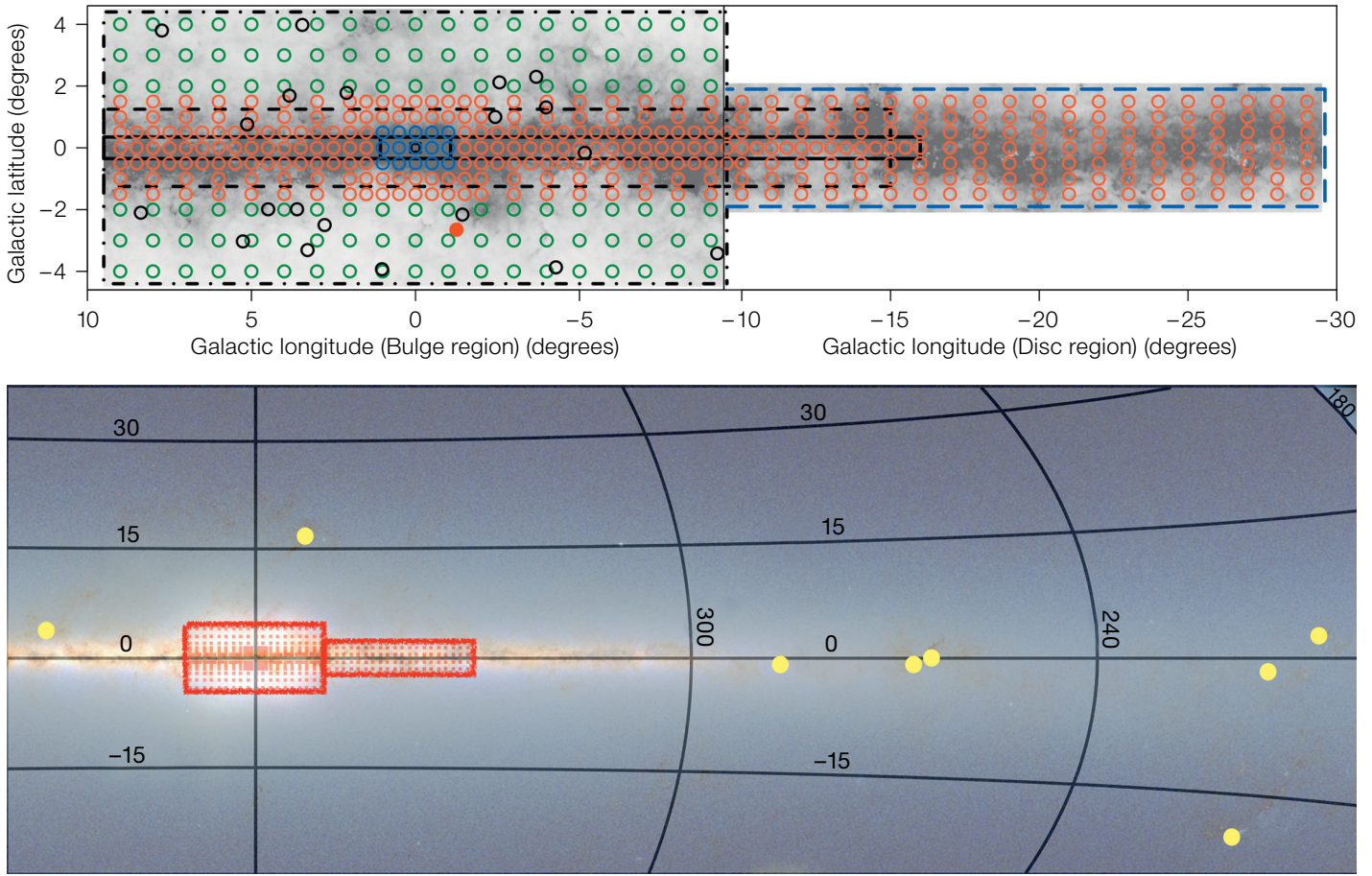

Figure 1. The top panel shows the layout of fields for the REDWAY survey. Lines mark the areas mapping the in-plane bar and inner disc (orange), boxy bulge (green), nuclear bulge (blue), bulge deep field (red), and inner galaxy clusters (black). The lower panel shows a zoomed-out view of the survey footprint showing in yellow circles the location of young clusters and star-forming regions to be mapped.

### The boxy/peanut bulge

The bulge region at $2 < |b| < 4$ degrees is the most favourable field in which to characterise the stellar population of the main bulge/bar itself without contamination from other nuclear components where other processes such as star formation might be at work, and with sufficient statistics to fully characterise the detailed chemical content and kinematics.

Recent optical surveys such as the Abundances and Radial velocity Galactic Origins Survey (ARGOS), the GIRAFFE Inner Bulge Survey (GIBS), and the Gaia–ESO Survey have unambiguously demonstrated that the bulge hosts at least two main components — a metal-poor one centred at [Fe/H] = –0.4 and a metal-rich one centred at [Fe/H] = +0.3. The two main bulge components have different kinematics and different spatial distributions. Recently, a study using the APOGEE data by Rojas-Arriagada et al. (2019) has shown that the magnesium-to-iron abundance of bulge stars is also bimodal, the separation coinciding with the separation in [Fe/H] abundance between the metal-poor and metal-rich populations. Therefore, a proper chemical separation of the two components will enable a cleaner characterisation of the kinematical and spatial properties of each of them.

The main issues that are currently unresolved are: (i) how the metal-poor component formed; and (ii) whether the large population of RR Lyrae stars in the bulge, tracing the component older than 10 Gyr, is associated with the metal-poor population or traces a third component. The high multiplexity of MOONS will allow us to perform a detailed analysis of the different populations for each line of sight, without the need to group stars to increase number statistics, therefore conserving the spatial resolution of the survey footprint. The MOONS REDWAY survey will produce a total sample of more than 102 000 stars distributed across the ~ 120 fields shown in Figure 1 in green at a SNR ~ 100 in the H band at R ~ 19 700.

### Bulge Deep Field

Despite the huge observational effort invested over the last decade in unveiling the star formation history of the Milky Way bulge, the missing key ingredient remains the accurate age of its stellar population. The scenario of an old (> 10 Gyr) bulge suggested by several photometric studies has been challenged by spectroscopic studies of main sequence turnoff stars that point towards the presence of a significant fraction of metal-rich stars as young as ~ 2–5 Gyr (Bensby et al., 2017). Furthermore, Haywood et al. (2016) have presented a study showing that the colour spread at the turnoff should be broader than observed, when accounting for the metallicity distribution of the bulge. They provide evidence suggesting that the correlation between age and metallicity can mimic the colour-magnitude diagram of a purely old population and showing that young stars are indeed necessary to reproduce the bulge colour-magnitude diagram. This discrepancy, and the possible explanation provided by Haywood et al. (2016), can only be settled by measuring individual ages for dwarf stars with stellar parameters and element abundances via high-resolution spectroscopy.

MOONS can observe several hundreds of main sequence turnoff stars in the Hubble Space Telescope bulge field known as the SWEEPS field $(l, b) = (1.25, -2.65)$, which has a baseline of proper motions covering over a decade under the WFC3 Galactic Bulge Treasury programme (Brown et al., 2009). This field (shown in Figure 1 as a red circle) will be used to anchor new astrometric catalogues covering the MOONS FOV (such as DECam, Gaia DR4, LSST, and dedicated programmes) to perform a proper-motion decontamination of the bulge turnoff sequence for target selection. The MOONS main sequence turnoff stars sample will reach a SNR > 50 in the $H$ band at $R \sim 19\,700$, thus providing radial velocities, stellar parameters and detailed abundances. A similar strategy will be applied to Baade's Window, which requires a shorter exposure time and suffers from fewer crowding problems. These two fields will generate the largest spectroscopic sample of turnoff bulge stars (at high spectral resolution), increasing the currently available sample by a factor of 10. This will allow for a robust measurement of the age distribution of the bulge and the investigation of vertical variations as well as enabling a comparison/validation of the photometric techniques applied to both fields. It will also allow for the first time the age-metallicity-orbital behaviour of a galactic bulge to be investigated on a star-by-star basis.

### Inner Milky Way Globular Clusters

MOONS offers a unique opportunity to build a homogeneous dataset of stellar abundances in each cluster, overcoming the limitation of low number statistics of previous studies, and comparing the properties of the globular cluster system with those of the field populations. The innermost Galactic regions probed by MOONS contain about one third of the whole Galactic globular cluster population. By observing tens of stars in most of these clusters, MOONS can provide the largest sample of detailed abundances of Galactic globular clusters, which will prove fundamental for all studies of Galactic stellar populations, and also for the study of globular cluster formation and early evolution (i.e., multiple populations).

The clusters to be targeted will be those for which spectroscopic observations are difficult owing to strong dust obscuration and a high density of targets (thus requiring high multiplexity over an arcminute-scale region). A total of 20 clusters will be sampled with dedicated pointings as shown in Figure 1 (black circles). In addition, fibres allocated to targets in newly discovered star clusters (for example from the VVV survey) will be included as part of the regular field survey footprint. Radial velocities, stellar parameters and abundances will be obtained for as many cluster members as possible. The number of targets will depend on the density and projected size of each cluster. However, MOONS is able to allocate tens of targets even in a very small cluster (such as NGC 6522). This is not currently possible with any other instrument in the near-infrared. High-signal-to-noise (SNR > 60 in the $H$ band) spectra will be obtained for all observed stars to derive accurate abundances down to a magnitude of $H_{AB} = 17.3$, spanning the red giant branch of even the most reddened clusters (at least 4 magnitudes below the tip of the red giant branch).

### Young Star Clusters

Large surveys like Gaia–ESO have yielded new results on the kinematic and dynamical properties of visible stars in (very) young clusters, such as complex kinematic structures and the puzzling discrepancy between the velocity dispersion of already formed stars and that of pre-stellar cores. Along with Gaia data, these results are revolutionising our understanding of cluster formation and early dynamical evolution.

The MOONS survey will complement surveys carried out at visible wavelengths by extensively mapping very young clusters and star-forming regions (Figure 1, bottom panel) that are partially or fully embedded and thus not accessible to previous studies, extending the critical range of masses, densities, and metallicities. The immediate goal is the determination of a) precise radial velocities of unbiased, statistically significant samples of cluster candidates, in order to derive the kinematics and dynamical status of the embedded and very low mass cluster populations; b) stellar parameters and accretion properties to fully characterise them; and c) metallicity and elemental abundances (sampling all nucleosynthetic channels) of the bright cluster members.

The survey will focus on regions where enough targets are available to exploit MOONS capabilities. It will prioritise observations in high-resolution mode down to $H_{AB} \sim 18.5$ and it will include faint and bright configurations in the same pointing. Bright stars will reach sufficiently high SNR (> 50) to derive elemental abundances, while for the fainter candidates only stellar parameters and radial velocities will be derived (SNR > 15–20). For both bright and faint candidate targets we will be able to derive accretion properties (the SNR will be higher in the emission lines).

### The Milky Way Satellites

The LMC and SMC are the largest and most massive satellites of the Milky Way. They are classified as irregular galaxies characterised by extended, still-ongoing star formation activity, as witnessed by the wide range of ages and metallicities of their stellar populations. On the other hand, the Sagittarius dwarf spheroidal galaxy is the most spectacular case of a Galactic satellite that is being disrupted by the Milky Way tidal field (Ibata et al.,1994), as witnessed by a two-arm tidal stream that has been traced across the entire sky (Majewski et al., 2003). The disruption of the Sagittarius dwarf spheroidal galaxy is contributing to the buildup of the Galactic halo in terms of dark matter, stars, and globular clusters.

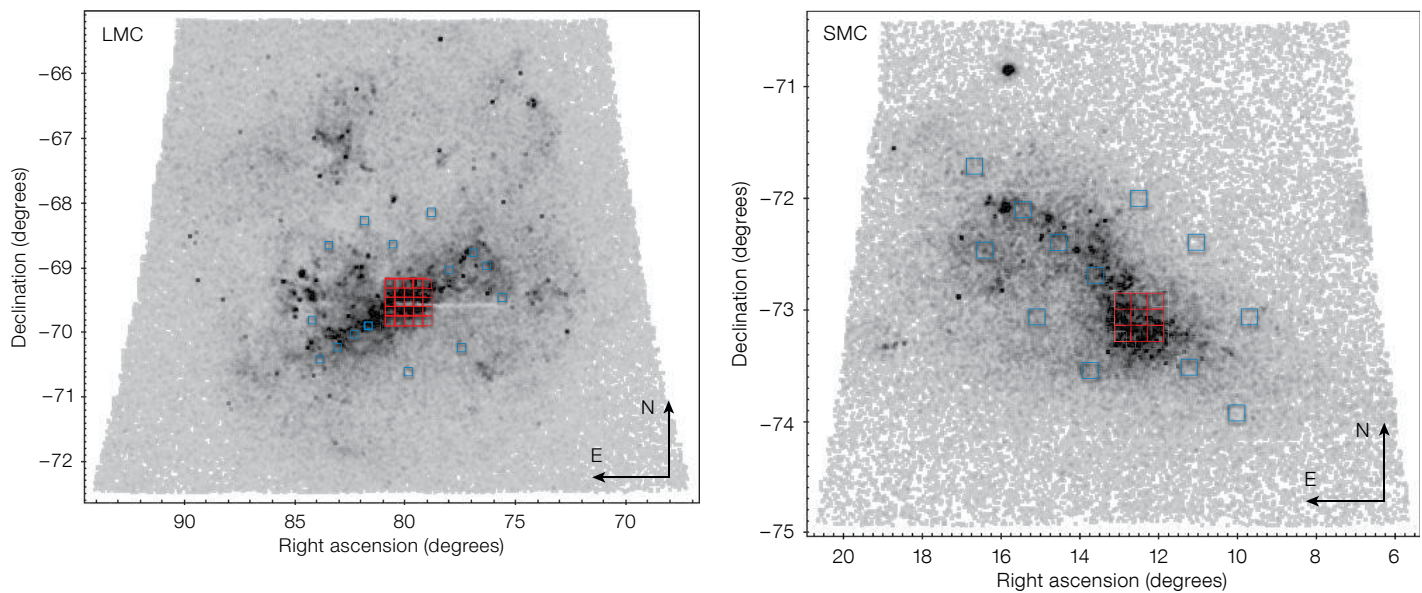

Figure 2. Map of the giant stars in the LMC (left) and the SMC (right). Red squares mark the first priority MOONS fields and blue squares show the second priority fields.

The multiplexing power of MOONS, coupled with its simultaneous wide spectral coverage and medium-high spectral resolution in the $I$ and $H$ bands, will allow us to dramatically increase the number of observed stars in the Magellanic Clouds and the Sagittarius dwarf spheroidal galaxy for which we have high quality spectra, thus providing an excellent characterisation of their stellar populations in terms of their chemistry and kinematics. A total of 116 fields covering both systems is planned to provide temperatures and surface gravities (from photometric and spectroscopic diagnostics), abundances of iron, α-elements (from optical and near-infrared spectra), CNO and possibly of a few other light elements (from near-infrared spectra only) and radial velocities.

Magellanic Clouds

The majority of studies providing information about the metallicity distribution in the Magellanic Clouds result from calcium triplet surveys (Carrera et al., 2011; Dobbie et al., 2014) of a few thousand stars. Our present knowledge of the chemical properties of the LMC is derived from optical observations collected with the ESO Fibre Large Array Multi Element Spectrograph (FLAMES) for evolved cool stars in the bright portion of the red giant branch ($V \sim 17$–17.5). Perhaps surprisingly, for a long time chemical information for SMC stars was limited to a few supergiant stars, located in some very young stellar clusters, while investigations of the stellar populations older than $\sim 2$ Gyr were lacking. Only very recently have Nidever et al. (2019) presented some chemical abundances of iron and α-elements from APOGEE spectra for luminous giant stars in the SMC.

However, the number of individual stars for which chemical abundances derived from high-resolution spectroscopy are available is insufficient to fully characterise the chemical enrichment history of the stellar populations of the Magellanic Clouds and to unveil possible gradients or differences in chemical enrichment in different regions of the galaxies. Similarly, existing observations of the kinematics of stars across the Magellanic Clouds are severely limited by the type of stellar populations, their spatial distribution, spectral resolution, and low number statistics. In particular, the central regions are poorly sampled, preventing us from being able to draw quantitative conclusions about the location of the gravitational centres.

The MOONS survey will produce a mosaic of 5 × 5 fields around the LMC centre (see Figure 2) to study the chemical composition and kinematics of the central bar and accurately identify the position of the LMC centre. It will also include about 17 fields located along the major axis of the bar and in the LMC disc (to investigate the chemical and kinematical properties in different LMC sub-structures) and four fields around the LMC centre, targeting the fainter, less evolved stellar population.

In the SMC, the survey will cover a mosaic of 3 × 3 fields around the SMC centre (see Figure 2) to study the chemical composition and kinematics of the densest region and accurately identify the position of the SMC centre, as well as about 13 fields along the SMC northeast extension to investigate the chemical and kinematical properties in regions associated with different sub-structures.

Additionally, a few tens of young/intermediate age stellar clusters and young clusters in binary systems will be also be observed in both galaxies. Studies of these systems can provide crucial information about the possible interactions suffered by the Magellanic Clouds in the past as well as on the mechanisms of cluster formation and evolution and their dependence on redshift and environment. Chemical abundance distributions and radial velocities will be derived for ~ 10–20 stars mapped in each cluster.

The combination of the derived abundances for stars (~ 35 000 stars in the LMC and ~ 18 000 in the SMC) down to $I_{AB}$ = 18 mag ($H_{AB}$ < 17.6 mag) from the MOONS survey with the star formation histories obtained from the VMC CMDs (Cioni et al., 2011), proper motions obtained from the VMC multi-epochs, and from Gaia (Gaia collaboration, Helmi et al., 2018), will provide a new picture of the structure of the Magellanic Clouds and the evolution of their stellar populations.

Sagittarius dwarf spheroidal

Several investigations of the chemical composition of the Sagittarius dwarf spheroidal galaxy are available in the literature but our knowledge of its chemistry is far from being complete or exhaustive. The metallicity distribution of the Sagittarius dwarf spheroidal galaxy is bimodal, with two main peaks at [Fe/H] = −0.5 and −1.5 dex, with an extended metal-poor tail (see for example, Bellazzini et al., 2008; Mucciarelli et al., 2017). However, the chemical composition and the distribution of the metal-poor population are poorly known as most of the available studies focus on the central region of the galaxy where the nuclear cluster M54 dominates over the Sagittarius field population in this metallicity regime (Bellazzini et al., 2008). For this reason, even if the Sagittarius dwarf spheroidal galaxy metallicity distribution is clearly dominated by its metal-rich, intermediate-age component, a complete and unbiased metallicity distribution is still lacking.

Several aspects of the chemical composition of the Sagittarius dwarf spheroidal galaxy remain unclear and large samples of medium-high resolution spectra are needed to solve some major open questions, namely: (i) its metallicity distribution, in particular the fraction of metal-poor stars (with [Fe/H] < −1 dex); (ii) the [α/Fe] abundance ratios that are ideal diagnostics for studying the contribution of Type II and Type Ia supernovae to the chemical enrichment of the system. The metal-rich component in the innermost part of the Sagittarius dwarf spheroidal galaxy shows sub-solar [α/Fe] and a knee located at [Fe/H] ~ −1.3 dex (de Boer, Belokurov, Koposov, 2015); and (iii) possible metallicity and radial velocity gradients along the main body of the galaxy. In particular, we stress that a milestone in the study of the chemical composition of the Sagittarius dwarf spheroidal galaxy will be to obtain an unbiased sample of stars in the galaxy, removing the contribution of M54.

The survey will cover a mosaic of 24 fields located within the inner 100 arcminutes of the central region of the Sagittarius dwarf spheroidal galaxy, outside the tidal radius of M54, to derive an unbiased metallicity distribution. It will also include about 16 fields outside of the core radius of the galaxy and along its major axis to investigate possible metallicity and kinematical gradients. The survey will map more than 15 000 stars down to the He clump, including about 5000 stars brighter than $I_{AB}$ < 18 mag, for the Sagittarius dwarf spheroidal galaxy centre (to obtain a robust metallicity distribution). In addition, two fields will be selected along the minor axis of the Sagittarius dwarf spheroidal galaxy to investigate possible metallicity and kinematical gradients. In the less dense, most external Sagittarius dwarf spheroidal galaxy regions, the survey will make efficient use of the MOONS fibres by observing additional Milky Way stars present along the line of sight. These observations will allow us to build up a database of MOONS spectra for several thousands of thick disc stars.

### Notes

[a] This is a conservative estimate of the number of fibres allocated to sky and is expected decrease once our sky subtraction techniques are optimised using on-sky data. Therefore, the total number of targets collected by the end of the survey is likely to be larger than given in this article.

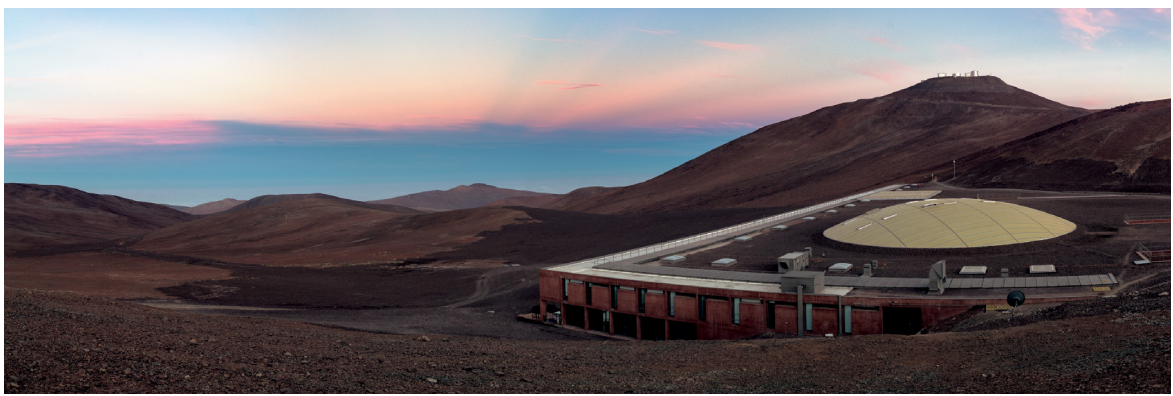

Early morning at Paranal Observatory. This image shows both the VLT and the Residencia, where staff and visiting astronomers and engineers stay during their visits to Paranal Observatory.